# Improved IKE Key Exchange Protocol Combined with Computer Security USB Key Device

Pak Myong-Suk, Jo Hyon-Chol, Jang Chung-Hyok

**Kim Il Sung** University,

Pyongyang, DPR of Korea

**Abstract:** In this paper we suggest improved IKE key exchange protocol combined with the Computer Security USB Key device to solve the problems in using IKE and IKE v2 protocol.

1. Introduction

The network layer virtual private network programs such as "strongSwan" and "Openswan" support both of the IKE and the IKEv2, but many networks still use IKE.

Unlike IKEv2, DoS(Denial of service) attack may happen in IKE protocol.(Dos attack for the DH calculation that happens when a lot of aggressive mode IKE requests having forged source IP addresses are received)[1]

In IKE/IKEv2, Man-in-the-middle attacks to SA payload and KE payload may happen, when user uses the electronic certificate distributed as the file format (eg. *p12), the authentication function of the user's certification can be dropped because of the electronic certificate keeping problem, so that the reliability of network communication can be decreased.[2,3]

## 2. Composition of the Computer Security USB Key device

To solve the problems in IKE/IKEv2, Computer Security USB Key device is used.

The Computer Security USB Key device is composed of CPU, NAND memory, power unit, USB connector.

NAND memory is divided into manager region where user can not read and write, virtual CD region where only reading is possible and the user's region where reading and writing are possible. Manager's region is divided into private key storing region, encrypt algorithm region and electronic certificate storing region.

In private key storing region of the manager's region the keys(or data that can create key) that can be used in security program or encrypt algorithm can be stored and device serial number for device uniqueness exists.

## 3. Improved IKE key exchange protocol combined with the Computer Security USB Key device and its Implementation

IKE $1^{st}$ phase security negotiation process in aggressive mode is used as an example for suggested method.

Initiator(i)                                         responder(r)

HDR, SA, KE, Ni, IDii, *UMi        →

← HDR, SA, KE, Nr, IDir, * UMr, CERT, SIG_R

HDR, *CERT, SIG_I            →

In negotiation process, * means that next payloads are encrypted and UMi and UMr are payloads suggested newly in this paper.

Improved IKE $1^{st}$ phase security negotiation process is as follows.

① Before the payload SA, KE, Ni and IDii are transmitted, the initiator should get the device serial number from the Computer Security USB Key device, makes UMi payload with device serial number and then encrypts it by using encryption key "key1" inside the device (key1 is same for all Computer Security USB Key device) to transmit to the responder.

If initiator can't get serial number, IKE $1^{st}$ phase security negotiation process is stopped.

② The responder recognizes the initiator as the legal user which didn't do DoS attack if encrypted UMi payload is decrypted successfully using encryption key "key1" of its Computer Security USB Key device and continues next stage.

In this stage too, the responder who doesn't have the Computer Security USB Key device can't take part in negotiation, so that the function of principal's identity authentication is raised to protect DoS attack from above two stages.

③ The responder works as the initiator to make UMr payload, generates signature using electronic certification kept in the Computer Security USB Key device of responder and encrypts CERT and SIG_R payload reflected electronic certification and signature as an encryption key(serial number of its Computer Security USB Key device) respectively to send to initiator. (In fact, CERT and SIG_R payload is transmitted as the plain text in aggressive mode. It is important to make these payload encrypt in order to raise the identity authentication function.)

④ After the initiator makes sure the responder's identity by decrypting encrypted UMr , CERT, SIG_R payloads, he makes signature by using electronic certification kept on initiator's Computer Security USB Key device and encrypts CERT and SIG_R with electronic certification and signature as an encryption key(serial number of its Computer Security USB Key device)respectively to send to responder.

⑤ The responder makes sure the initiator's identity by decrypting encrypted CERT, SIG_I payloads respectively.

Encrypt algorithm used in improved IKE $1^{st}$ phase security negotiation process can be done by using encrypt algorithm kept on the Computer Security USB Key device.

The type of payload(UMi, UMr) including the information of Computer Security USB Key device in IPsec(ex, ipsec-tools-0.6.5-9.el5.src.rpm)security program can be defined as follows.

```
#define ISAKMP_NPTYPE_DEV     55     /*device information */
```

And the structure of this payload can be defined as follows.

```
typedef struct _devinfo_t_ {
unsigned char dev_serial[7]; /* the serial number of USB key device */
} devinfo_t;
```

## Result analysis and Conclusion

The table 1 shows comparison between previous IKE protocol and proposed method in this paper.

Table 1.  Comparison of security performance

| Protocol name | SA, KE payload protection | CERT, SIG payload protection | DoS attack prevention | extensibility | Certificate storage |
|---|---|---|---|---|---|
| IKE (IKE 1st phase aggressive mode) | × | × | × | × | file |
| Improved IKE (IKE 1st phase aggressive mode) | ○ | ○ | ○ | IKEv2 | Computer Security USB Key Device |

Where , × indicates no support , ○ indicates support.

Improved IKE key exchange protocol uses default SA and default KE payload, but encryption algorithm and key used in IKE 1st phase security negotiation process are encryption algorithm and key kept in computer security USB key device, so that man-in-the-middle attacks to SA payload and KE payload are prevented.

And DOS attack is prevented by using UMi, UMr payload with device information of user, electronic certificate and signature payload are encrypted in computer security USB key device and exported, so that reliability for identify authentication is raised.

The suggested method can be also applied in IKE v2 protocol.